\documentclass[pre,aps,preprint,amsmath,amssymb,showpacs]{revtex4}
\usepackage{graphicx}
\usepackage{epsfig}
\usepackage{epstopdf}
\include{graphics}

\DeclareMathOperator{\sech}{sech}

\usepackage{amssymb,graphicx,amsthm}
\usepackage{soul}

\usepackage[most]{tcolorbox}
\newtcolorbox{highlighted}{colback=yellow,coltext=red,breakable}

\begin{document}

\title{Collision-induced amplitude dynamics of pulses in linear waveguides with the generic nonlinear loss}

\author{Quan M. Nguyen$^{1}$}

\affiliation{$^{1}$Department of Mathematics, International University, 
Vietnam National University-HCMC, Ho Chi Minh City, Vietnam}

\date{\today}

\begin{abstract}

We study the effects of the generic weak nonlinear loss on fast two-pulse interactions in linear waveguides. The colliding pulses are described by a system of coupled Schr\"odinger equations with a purely nonlinear coupling in the presence of the weak $(2m+1)-$order of nonlinear loss, for any $m \geq 1$. We derive the analytic expression for the collision-induced amplitude shift in a fast two-pulse interaction. The analytic calculations are based on a generalization of the perturbation technique for calculating the effects of weak perturbations on fast collisions between solitons of the nonlinear Schr\"odinger equation. The theoretical predictions are confirmed by the numerical simulations with the full propagation model of coupled Schr\"odinger equations. 

\end{abstract}
\maketitle

\section{Introduction}
\label{Introduction}
 
Linear and nonlinear waves are widely studied and used in a variety of physical applications \cite{Whitham74, Boyd2008, Keller1979, Eckhoff84, Chen2016, Johnson2000}. In linear optical waveguides, the dynamics of pulses can be affected by nonlinear loss \cite{Boyd2008, Agrawal2007a}. 
Nonlinear loss arises in optical waveguides due to 
multiphoton absorption or gain/loss saturation \cite{Boyd2008, Agrawal2007a}. More specifically, the $(2m+1)-$order of loss can be a result of $(m+1)$-photon absorption in silicon waveguides \cite{Boyd2008}. The $M$-photon absorption with $2\le M \le 5$ has been 
the subject of intensive theoretical and experimental research 
in recent years due to a wide variety of potential applications,  
including lasing, material processing, and optical data storage, etc. 
\cite{Boyd2008, Yoshino2003, Agrawal2007a, Husko2013a, Husko2013b, Zheng2013, Loon2018, PNH2017}. Therefore, it is very important to study the impact of nonlinear loss on the propagation and dynamics of pulses in linear and nonlinear waveguides. In nonlinear waveguides, the balance between the dispersion and the nonlinearity can form optical solitons \cite{Boyd2008}. The impacts of nonlinear loss on optical solitons were studied in several earlier papers, e.g. Refs. \cite{Agrawal2007a,Husko2013a,PCG2003, CP2005, PNC2010, PNG2014}. In particular, the expressions for collision-induced amplitude shift in a fast two-soliton collision in the presence of weak cubic loss and in the presence of the weakly generic nonlinear loss were set up and verified by numerical simulations in  Refs. \cite{PNC2010} and \cite{PNG2014}, respectively. In these studies, the calculations were based on the shape-preserving and stability properties of the solitons. For these reasons, it was often
claimed that conclusions drawn from analysis of soliton collisions cannot be applied to collisions between pulses of
weakly perturbed linear systems, where the pulse shapes are not preserved during the collisions \cite{Boyd2008,PCG2003, CP2005, PNC2010, PNG2014}.

Recently, in Ref. \cite{PNH2017}, the authors partially showed that this point of view on fast collisions between pulses that are not shape preserved, as mentioned above, is erroneous. More specifically, the authors have found an expression for collision-induced amplitude shift in a fast collision
between two pulses in linear waveguides in the presence of the weak cubic loss, that is, in a specific case of $m=1$, which is described by a coupled system of non-solitonic equations. In addition, in this work, the authors also demonstrated that pulses in linear waveguides with weak cubic loss exhibit soliton-like behavior in fast two-pulse collisions. This made an interesting connection between the collision of two
quasi-linear pulses with the one of two solitons of the nonlinear Schr\"odinger equation. However, so far, a comprehensive theoretical study of the effects of the generic $(2m+1)-$order of loss, for any $m \geq 1$, on two-pulse interactions in linear waveguides is still an open problem.

In the current paper, we address this important and interesting problem. First, we derive the equation for amplitude dynamics of a single pulse in the presence of the generic weak nonlinear loss, i.e., the $(2m+1)-$order of loss, for any $m \geq 1$. This can be calculated in
a straightforward manner by implementing the standard adiabatic
perturbation theory. Second, we derive the collision-induced amplitude shift in a fast collision
between two pulses in the presence of the generic weak nonlinear loss. The calculations are based on deriving and integrating the partial differential equation for the collision-induced change in the envelope of pulse at the leading order of the perturbative calculation. We show that the nonlinear loss also strongly affects the collisions of pulses, by causing an additional decrease
of pulses amplitudes. Finally, we validate our theoretical calculations by numerical simulations with the propagation model for $m=2$ and $m=3$. The calculations of the collision-induced amplitude shift in the current paper 
are based on an extension of the perturbation technique, developed in Refs. 
\cite{PCG2003, CP2005, PNC2010, PNG2014} for calculating the effects of weak perturbations 
on fast collisions between solitons of the nonlinear Schr\"odinger equations and between pulses of coupled partial differential equations (PDEs) with the weak cubic loss in \cite{PNH2017}.

The rest of the paper is organized as follows. In section \ref{collision}, we introduce the propagation model and derive the equations for amplitude dynamics of a single pulse and for collision-induced amplitude dynamics. In section \ref{simul}, we validate the theoretical calculations by simulations. Section \ref{concl} is reserved for conclusions.

\section{Pulse interaction in linear waveguides with the generic weak nonlinear loss}
\label{collision} 

We consider fast collisions between two optical pulses in linear waveguides in the presence of the weak $(2m+1)-$order of the nonlinear loss for $m \geq 1$. The propagation equations can be given by the following coupled Schr\"odinger equations with a purely nonlinear coupling in the presence of the nonlinear loss \cite{Agrawal2007a, PNH2017, PNC2010, PNG2014}: 
\begin{eqnarray}&&
\!\!\!\!\!\!\!
i\partial_{z}\psi_{1} - \mbox{sgn}(\tilde\beta_{2})\partial_{t}^{2}\psi_{1} = -i\epsilon_{2m+1}|\psi_{1}|^{2m}\psi_{1} - i\epsilon_{2m+1}\sum\limits_{k = 1}^{m} {b_{k}|\psi_{2}|^{2k}|\psi_{1}|^{2(m-k)}\psi_{1} } ,
\nonumber \\&&
\!\!\!\!\!\!\!
i\partial_{z}\psi_{2} + id_{1}\partial_{t}\psi_{2} - \mbox{sgn}(\tilde\beta_{2})\partial_{t}^{2}\psi_{2} = -i\epsilon_{2m+1}|\psi_{2}|^{2m}\psi_{2} 
- i\epsilon_{2m+1}\sum\limits_{k = 1}^{m} {b_{k}|\psi_{1}|^{2k}|\psi_{2}|^{2(m-k)}\psi_{2} },  
\!\!\!\!\!\!\!\!
\nonumber \\&&
\label{coll1}
\end{eqnarray}           
where $b_{k}=\frac{m!(m+1)!}{(k!)^{2}(m+1-k)!(m-k)!}$, $\psi_{1}$ and $\psi_{2}$ are proportional to the envelopes of the electric fields of the pulses, 
$z$ is the (normalized) propagation distance, and $t$ is the time \cite{Dimensions1}. 
In Eq. (\ref{coll1}), $\tilde\beta_{2}$ is the second-order dispersion coefficient, $d_{1}$ is the group velocity coefficient and 
$\epsilon_{2m+1}$ is the generic nonlinear loss coefficient in the weak $(2m+1)$-order of loss, $0 <\epsilon_{2m+1} \ll 1$, for $m\geq 1$. The first and second terms on the right hand side of Eq. (\ref{coll1}) describe intra-pulse and inter-pulse effects due to the $(2m+1)$-order of loss. It is worthy to note that the perturbed coupled propagation model (\ref{coll1}) is based on the assumption 
that the effects of cubic nonlinearity, i.e., Kerr nonlinearity, can be neglected. 
This assumption was successfully used in earlier experimental and 
theoretical works, see e.g., Refs. \cite{Perry97,Liang2005,Cohen2005b,Cohen2004}.

We consider an initial pulse $\psi_{j}(t,0)$ which its tails exponentially decay. We assume that the pulses can be characterized by 
initial amplitudes $A_{j}(0)$, initial widths $W_{j0}$, initial positions $y_{j0}$, 
and initial phases $\alpha_{j0}$, 
such that $\psi_{j}(t,0)$ can be expressed in the 
general form  
\begin{eqnarray} &&
\psi_{j}(t,0)=A_{j}(0)f_{j}\left[(t-y_{j0})/W_{j0}\right]\exp(i\alpha_{j0}),  
\label{ICA}
\end{eqnarray} 
where $f_{j}(y)$ are real-valued functions of $y$, and $j=1,2$. For example, one can use $f_{j}(y)=\exp(-y^2/2)$ for Gaussian pulses or $f_{j}(y)=\sech(y)$ for hyperbolic secant pulses.

First, we study the amplitude dynamics of a single pulse in the presence of the generic $(2m+1)-$order of nonlinear loss described by the following equation:
\begin{eqnarray} &&
\!\!\!\!\!\!\!
i\partial_{z}\psi_{j} + id_{1}\partial_{t}\psi_{j} - \mbox{sgn}(\tilde\beta_{2})\partial_{t}^{2}\psi_{j} = - i\epsilon_{2m+1}|\psi_{j}|^{2m}\psi_{j}.
\!\!\!\!\!\!\!
\label{single1}
\end{eqnarray}
By deriving the energy balance of Eq. (\ref{single1}), it yields 
\begin{eqnarray}&&
\!\!\!\!\!\!\!\!\!\!\!\!
\partial_{z}\int_{-\infty}^{\infty} \!\!\!\!\!\!|\psi_{j}(t,z)|^{2}dt\!= -2\epsilon_{2m+1}\int_{-\infty}^{\infty} \!\!\!\!\!\!|\psi_{j}(t,z)|^{2m+2}dt.
\label{single2}
\end{eqnarray}       
We express the approximate solution of the propagation equation (\ref{single1}) as 
$\psi_{j}(t,z)=A_{j}(z)\tilde\psi_{j}(t,z)$,
where $A_{j}(z)$ is the amplitude parameter and $\tilde\psi_{j}(t,z)$ is the solution 
of the propagation equation in the absence of $(2m+1)-$order of loss
with initial amplitude $A_{j}(0)=1$:
\begin{eqnarray}&&
\!\!\!\!\!\!\!\!\!\!\!\!
\tilde\psi_{j}(t,z)=\tilde\Psi_{j0}(t,z)\exp[i\chi_{j0}(t,z)],
\label{pulse1}
\end{eqnarray}       
where $\tilde\Psi_{j0}(t,z)$ and $\chi_{j0}(t,z)$ are real-valued. Substituting the relation for 
$\psi_{j}(t,z)$ into Eq. (\ref{single2}), it implies 
\begin{eqnarray} &&
\!\!\!\!\!\!\!
\frac{d}{dz}\left[ I_{2j}(z)A_{j}^{2}(z) \right] = - 2\epsilon_{2m+1}I_{2m+2,j}(z)A_{j}^{2m+2}(z),
\!\!\!\!\!\!\!
\label{ODE1}
\end{eqnarray}
where $I_{2j}(z)=\int_{-\infty}^{\infty} |\tilde\psi_{j}(t,z)|^{2}dt=I_{2j}(0)$ by the conservation of energy for the unperturbed solution $\tilde\psi_{j}(t,z)$, and $I_{2m+2,j}(z)=\int_{-\infty}^{\infty} |\tilde\psi_{j}(t,z)|^{2m+2}dt$. Integrating the differential equation (\ref{ODE1}) by a change of variable of $S_{j}(z)=A_{j}^2(z)$, one can obtain the equation for amplitude dynamics of a single pulse: 
\begin{eqnarray} &&
\!\!\!\!\!\!\!
A_{j}(z)= \frac{A_{j}(0)}{ \left[1 + 2m\epsilon_{2m+1}I_{2j}^{-1}(0)\tilde I_{2m+2,j}(0,z)A_{j}^{2m}(0)\right]^{1/(2m)} } ,
\!\!\!\!\!\!\!
\label{Amplitude1}
\end{eqnarray}
where $\tilde I_{2m+2,j}(0,z)=\int_{0}^{z} {I_{2m + 2,j}\left( {z'} \right)dz'}$.

Second, we calculate the collision-induced amplitude dynamics in a fast collision between two pulses with generic shapes 
and with tails that exhibit exponential decay. We consider a complete collision, i.e., the two pulses are well separated at the initial distance $z=0$ 
and at the final distance $z=z_{f}$. We define the collision length $\Delta z_{c}$, 
which is the distance along which the envelopes of the colliding pulses overlap, 
by $\Delta z_{c}=W_{0}/|d_{1}|$, where for simplicity we assume $W_{10}=W_{20}=W_{0}=\mathcal{O}(1)$. The condition for a fast collision is $z_{D} \gg \Delta z_{c}$, where $z_D=W_{0}^2/2$ is the dispersion length. That is, $W_{0}|d_{1}| \gg 1$ \cite{PNH2017}. Therefore, with the assumption of $W_{0}=\mathcal{O}(1)$, this is equivalent to $|d_{1}| \gg 1$, which allows us to use the two small parameters $\epsilon_{2m+1}$ and $1/|d_1|$ for the perturbative calculations. These conditions are realistic in optical fiber transmission systems, see, for example, \cite{PNH2017} and references therein, for an experimental setup with $W_0=2$ and $d_1 \gg 1$. In an analogy with the perturbative calculation approach in \cite{PNH2017, CP2005, PNC2010, PNG2014},  we look for a solution of Eq. (\ref{coll1}) in the form 
\begin{eqnarray}&&
\!\!\!\!\!\!\!
\psi_{j}(t,z)=\psi_{j0}(t,z)+\phi_{j}(t,z), 
\label{coll2}
\end{eqnarray}       
where $j=1,2$, $\psi_{j0}$ are the solutions of Eq. (\ref{coll1}) without 
the inter-pulse interaction terms, and $\phi_{j}$ describe corrections to 
$\psi_{j0}$ due to inter-pulse interaction. That is, $\psi_{10}$ and $\psi_{20}$ satisfy 
\begin{eqnarray}&&
\!\!\!\!\!\!\!
i\partial_z\psi_{10}\!-\!\mbox{sgn}(\tilde\beta_{2})\partial_{t}^{2}\psi_{10} =
-\!i\epsilon_{2m+1}|\psi_{10}|^{2m}\psi_{10},
\!\!\!\!\!\!\!\!
\label{coll3}
\end{eqnarray}         
and
\begin{eqnarray}&&
\!\!\!\!\!\!\!
i\partial_z\psi_{20}+id_{1}\partial_{t}\psi _{20}
-\mbox{sgn}(\tilde\beta_{2})\partial_{t}^2\psi_{20} = -i\epsilon_{2m+1}|\psi_{20}|^{2m}\psi_{20},
\label{coll4}
\end{eqnarray}   
where the initial conditions are $\psi_{j0}(t,0)=\psi_j(t,0)$ given by Eq. (\ref{ICA}), for $j=1,2$. We substitute relation (\ref{coll2}) into (\ref{coll1}) 
and use Eqs. (\ref{coll3}) and  (\ref{coll4}) 
to obtain equations for the $\phi_{j}$. Taking into account only leading-order effects of the collision, 
we can neglect terms containing $\phi_{j}$ on the right hand side 
of the resulting equation. One can therefore obtain the equation for $\phi_1$: 
\begin{equation}
i\partial_z\phi_{1}-\mbox{sgn}(\tilde\beta_{2})\partial_{t}^{2}\phi_{1}=
- i\epsilon_{2m+1}\sum\limits_{k = 1}^{m} {b_{k}|\psi_{20}|^{2k}|\psi_{10}|^{2(m-k)}\psi_{10} }.
\label{coll5} 
\end{equation}      
Let $\psi_{j0}(t,z)=\Psi_{j0}(t,z)\exp[i\chi_{10}(t,z)]$ and $\phi_{1}(t,z)=\Phi_{1}(t,z)\exp[i\chi_{10}(t,z)]$, where $\chi_{10}(t,z)$ is defined from Eq. (\ref{pulse1}). We substitute $\psi_{j0}(t,z)$ and $\phi_{1}(t,z)$ into Eq. (\ref{coll5}).
This substitution yields the following equation for $\Phi_{1}$: 
\begin{eqnarray} &&
i\partial_{z}\Phi_{1} - \left(\partial_{z}\chi_{10}\right)\Phi_{1}
-\mbox{sgn}(\tilde\beta_{2})\left[ 
\partial_{t}^{2}\Phi_{1}
+2i\left(\partial_{t}\chi_{10}\right)\partial_{t}\Phi_{1}
\right. 
\nonumber \\&&
\left. 
+i\left(\partial_{t}^{2}\chi_{10}\right)\Phi_{1}
-\left(\partial_{t}\chi_{10}\right)^2\Phi_{1}
\right] =
- i\epsilon_{2m+1}\sum\limits_{k = 1}^{m} {b_{k}\Psi_{20}^{2k}\Psi_{10}^{2(m-k)+1}}.
\label{coll6}
\end{eqnarray} 
Since the collision length $\Delta z_{c}$ is of order 
$1/|d_{1}|$, therefore, on the left-hand side of Eq. (\ref{coll6}), the only term $i\partial_{z}\Phi_{1}$, which contains the fast rate of change of $\Phi_{1}$ with respect to $z$ along the fiber, is of order 
$|d_{1}| \times \mathcal{O}(\Phi_{1})$ and other terms are of order $\mathcal{O}(\Phi_{1})$. In Eq. (\ref{coll6}), equating the leading orders of the left-hand side, which is of order $|d_{1}| \times \mathcal{O}(\Phi_{1})$, and of the right-hand side, which is of order $\mathcal{O}(\epsilon_{2m+1})$, this yields that $\Phi_{1}$ 
is of order $\epsilon_{2m+1}/|d_{1}|$. Therefore, in the leading order of the perturbative calculation, 
the equation for the collision-induced change in the envelope of pulse 1 is
\begin{eqnarray} &&
\partial_{z} \Phi_{1}(t,z)=- \epsilon_{2m+1}\sum\limits_{k = 1}^{m} {b_{k}\Psi_{20}^{2k}\Psi_{10}^{2(m-k)+1}}.
\label{coll7}
\end{eqnarray}
Let $z_c$ be the collision distance, which is the distance at which the maxima of $|\psi_{j}(t,z)|$ coincide. Thus, the fast collision takes place in the small interval $[z_{c}-\Delta z_{c}, z_{c}+\Delta z_{c}]$. Let $\Delta\Phi_{1}(t,z_{c}) = \Phi_1 \left( {t,z_{c} + \Delta z_{c}} \right) - \Phi_1 \left( {t, z_{c} - \Delta z_{c}} \right)$ be the {\it net} collision-induced change in the envelope of pulse 1. We substitute $\Psi_{j0}(t,z)=A_{j}(z)\tilde \Psi_{j0}(t,z)$ into Eq. (\ref{coll7}) and integrate with respect to $z$ over the interval $[z_{c}-\Delta z_{c}, z_{c}+\Delta z_{c}]$, we have
\begin{eqnarray} &&
\Delta\Phi_{1}(t,z_{c})=- \epsilon_{2m+1}\sum\limits_{k = 1}^{m}
{b_{k}J_{k,m}},
\label{coll8}
\end{eqnarray}
where $J_{k,m}=\int_{z_{c} - \Delta z_{c}}^{z_{c} + \Delta z_{c}} {A_{2}^{2k}(z')A_{1}^{2(m-k)+1}(z')\tilde \Psi_{20}^{2k}(t,z') \tilde \Psi_{10}^{2(m-k)+1}(t,z')dz'}$. To calculate $J_{k,m}$, we note that in the integrand of $J_{k,m}$, there is {\it only} one factor $\tilde\Psi_{20}(t,z')$ that contains the
dependence on the fast-changing variable $y=t-y_{20}-d_{1}z'$. Therefore, from Eq. (\ref{coll8}), we obtain the following approximation:
\begin{eqnarray} &&
\Delta \Phi_{1}(t,z_{c})=- \epsilon_{2m+1}\sum\limits_{k = 1}^{m}
{b_{k}A_{2}^{2k}(z_{c}^{-})A_{1}^{2(m-k)+1}(z_{c}^{-})\tilde \Psi_{10}^{2(m-k)+1}(t,z_{c})L_{k,m}},
\nonumber \\&&
\label{coll9}
\end{eqnarray}
where $L_{k,m}=\int_{z_{c}-\Delta z_{c}}^{z_{c}+\Delta z_{c}} {\tilde \Psi_{20}^{2k}(t,z')dz'}$ and $A_{j}(z_{c}^{-})$ denotes the limit from the left of $A_{j}$ at $z_{c}$. In calculating the integral $L_{k,m}$ one can take into account only the fast dependence 
of $\tilde\Psi_{20}$ on $z$, i.e., the $z$ dependence that is contained in factors of the form 
$y=t-y_{20}-d_{1}z$ and approximate other slow varying terms of $\tilde\Psi_{20}$ by their values at $z_c$. Denoting this approximation of 
$\tilde\Psi_{20}(t,z)$ by $\bar\Psi_{20}(y,z_{c})$, one can obtain  
\begin{eqnarray} &&
\Delta\Phi_{1}(t,z_{c}) = - \epsilon_{2m+1}\sum\limits_{k = 1}^{m}
{b_{k}A_{2}^{2k}(z_{c}^{-})A_{1}^{2(m-k)+1}(z_{c}^{-})\tilde \Psi_{10}^{2(m-k)+1}(t,z_{c})M_{k,m}},
\nonumber \\&&
\label{coll10}
\end{eqnarray}
where $M_{k,m}=\int_{z_{c}-\Delta z_{c}}^{z_{c}+\Delta z_{c}} {\bar \Psi_{20}^{2k}(t-y_{20}-d_{1}z',z_{c})dz'}$.
Since the integrand of $M_{k,m}$ is sharply peaked at a small interval about $z_c$, one can extend the limits of the integral to $-\infty$ and $\infty$ and change the integration variable from $z'$ to $y=t-y_{20}-d_{1}z'$ then obtain $M_{k,m}=\dfrac{1}{|d_{1}|}M_{k,m}^{'}$, where $M_{k,m}^{'}= \int_{-\infty}^{\infty} {\bar \Psi_{20}^{2k}(y,z_{c})dy}$. Based on Eq. (\ref{coll10}) and on the following relation between the net collision-induced change in the envelope and the collision-induced amplitude: $\Delta A_{1}^{(c)} \int_{-\infty}^{\infty} {\tilde \Psi_{10}^{2}(t,z_{c}) dt}= \int_{-\infty}^{\infty} {\Delta \Phi_{1}(t,z_{c}) \tilde \Psi_{10}(t,z_{c})dt}$ (see Eq. (12) in \cite{PNH2017}), one can derive the expression for the collision-induced amplitude shift of pulse 1:
\begin{eqnarray} &&
\Delta A_{1}^{(c)}=- \frac{\epsilon_{2m+1}}{|d_{1}|}\sum\limits_{k = 1}^{m}
{b_{k}A_{2}^{2k}(z_{c}^{-})A_{1}^{2(m-k)+1}(z_{c}^{-})N_{k,m}M_{k,m}^{'}},
\label{coll11}
\end{eqnarray}
where $N_{k,m}=\dfrac{\int_{-\infty}^{\infty} {\tilde \Psi_{10}^{2(m-k)+2}(t,z_{c})}dt}{\int_{-\infty}^{\infty} {\tilde \Psi_{10}^{2}(t,z_{c})}dt}$, for $1\le k \le m$. We emphasize that in the specific case of $m=1$, Eq. (\ref{coll11}) becomes Eq. (13) in \cite{PNH2017}. Also, it is worthy to remark that the analytic expression for $\Delta A_{1}^{(c)}$ in Eq. (\ref{coll11}) is independent of the pulse shapes of the colliding pulses.

\section{Numerical simulations}
\label{simul}

In this section, we shall validate Eq. (\ref{coll11}) by numerical simulations with the coupled PDEs of Eq. (\ref{coll1}). As a concrete example, we demonstrate the numerical simulations for a collision of two Gaussian pulses in the presence of the quintic loss ($m=2$) and septic loss ($m=3$). Equation (\ref{coll1}) is numerically integrated by implementing the split-step Fourier method with periodic boundary
conditions \cite{Hardin73, Yang2010, Glowinski2016}

The initial envelopes of the Gaussian pulses are $
\psi_{j}(t,0)=A_{j}(0)
\exp[-(t-y_{j0})^{2}/(2W_{j0}^{2})+i\alpha_{j0}]$, for $j=1,2$. Therefore, one obtains
\begin{eqnarray} &&
\tilde \Psi_{10}(t,z)=\frac{W_{10}}{(W_{10}^{4}+4 z^{2})^{1/4}}
\exp\left[\frac{-W_{10}^{2}(t-y_{10})^{2}}{2(W_{10}^{4}+4 z^{2})}\right],
\label{simul1a}
\end{eqnarray}
and
\begin{eqnarray} &&
\tilde \Psi_{20}(t,z)=\frac{W_{20}}{(W_{20}^{4}+4 z^{2})^{1/4}}
\exp\left[\frac{-W_{20}^{2}(t-d_{1}z-y_{20})^{2}}{2(W_{20}^{4}+4 z^{2})}\right]. 
\label{simul1b}
\end{eqnarray}
Therefore, in the calculations for Eqs. (\ref{coll10}) and (\ref{coll11}), the approximation of $\tilde\Psi_{20}(t,z)$ is 

\begin{eqnarray} &&
\bar\Psi_{20}(y,z_{c})=\frac{W_{20}}{(W_{20}^{4}+4 z_{c}^{2})^{1/4}}
\exp\left[\frac{-W_{20}^{2}y^{2}}{2(W_{20}^{4}+4 z_{c}^{2})}\right].
\label{simul1c}
\end{eqnarray}
Equations (\ref{simul1a}), (\ref{simul1b}), and (\ref{simul1c}) completely determine the theoretical prediction for $\Delta A_{1}^{(c)}$ in Eq. (\ref{coll11}).

We now demonstrate the method for the numerical measurement of $\Delta A_{1}^{(c)(num)}$. We first numerically calculate $A_{1}(z_{f})$ from the simulations with the full coupled PDEs of Eq. (\ref{coll1}). Note that that the colliding pulses are endured by the two processes: the self-amplitude shift during the propagation, which is theoretically described by Eq. (\ref{ODE1}) or Eq. (\ref{Amplitude1}), and the collision-induced amplitude shift lasting in the collision interval $[z_{c}-\Delta z_{c}, z_c+\Delta z_{c}]$, which is $\Delta A_{1}^{(c)(num)}$. Therefore, one can take into account the self-amplitude shift by calculating $A_{1}({z_{c}^{-}})$ from $A_{1}(0)$ thanks to Eq. (\ref{Amplitude1}) and by calculating $A_{1}({z_{c}^{+}})$ from $A_1(z_f)$:
\begin{eqnarray} &&
\!\!\!\!\!\!\!
A_{1}({z_{c}^{+}})= \frac{A_{1}(z_f)}{ \left[1 - 2m\epsilon_{2m+1}I_{2j}^{-1}(0)\tilde I_{2m+2,j}(z_c,z)A_{j}^{2m}(z_f)\right]^{1/(2m)} } ,
\!\!\!\!\!\!\!
\label{simul2a}
\end{eqnarray}
where $\tilde I_{2m+2,j}(z_c,z_f)=\int_{z_c}^{z_f} {I_{2m + 2,j}\left( {z'} \right)dz'}$. 
Finally, we measure numerical value of $\Delta A_{1}^{(c)(num)}$ by the difference between $A_{1}({z_{c}^{+}})$ and $A_{1}({z_{c}^{-}})$:
\begin{eqnarray} &&
\Delta A_{1}^{(c)(num)}=A_{1}({z_{c}^{+}}) - A_{1}({z_{c}^{-}}).
\label{simul2}
\end{eqnarray}

\begin{figure}[ptb]
\begin{tabular}{cc}
\epsfxsize=13cm  \epsffile{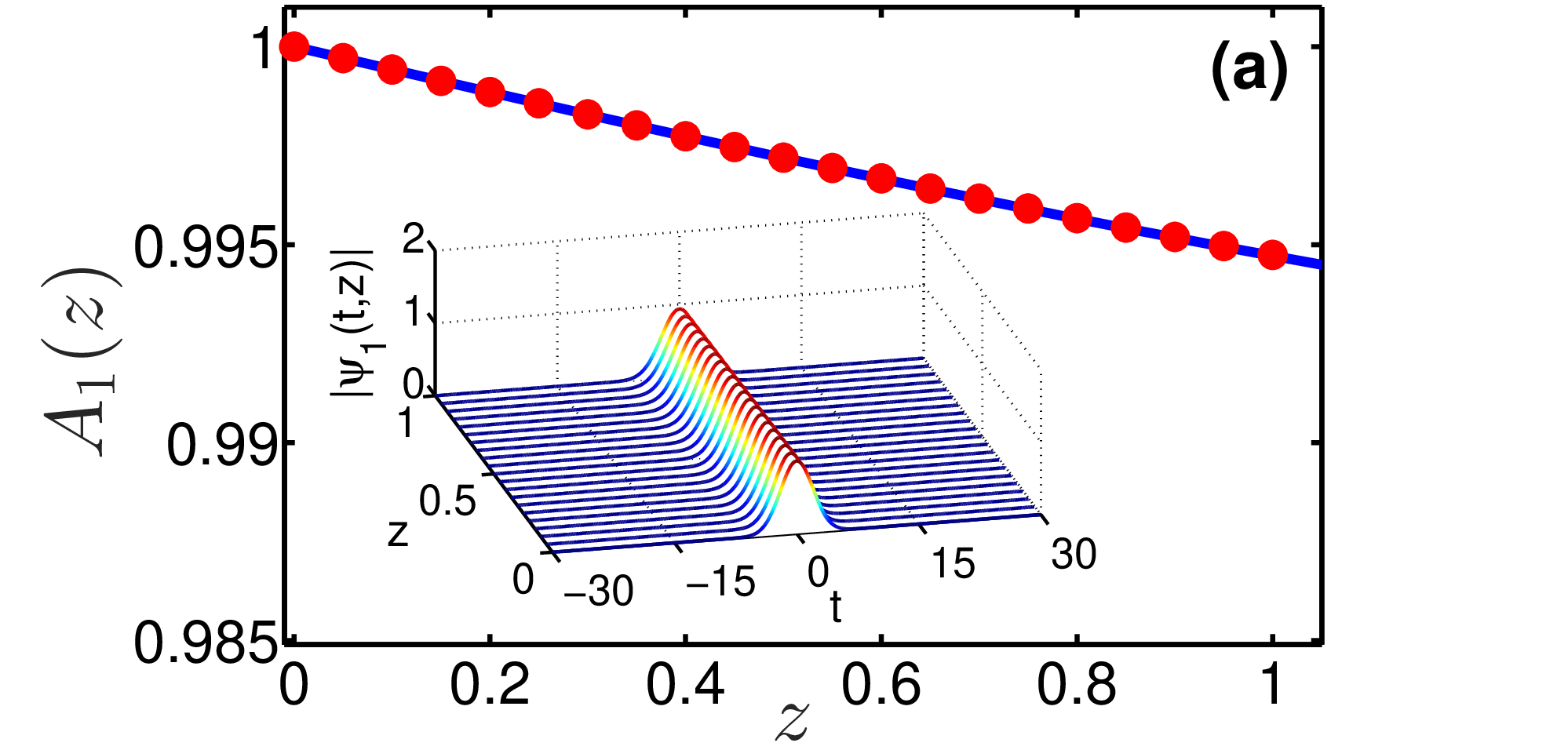} \\
\epsfxsize=13cm  \epsffile{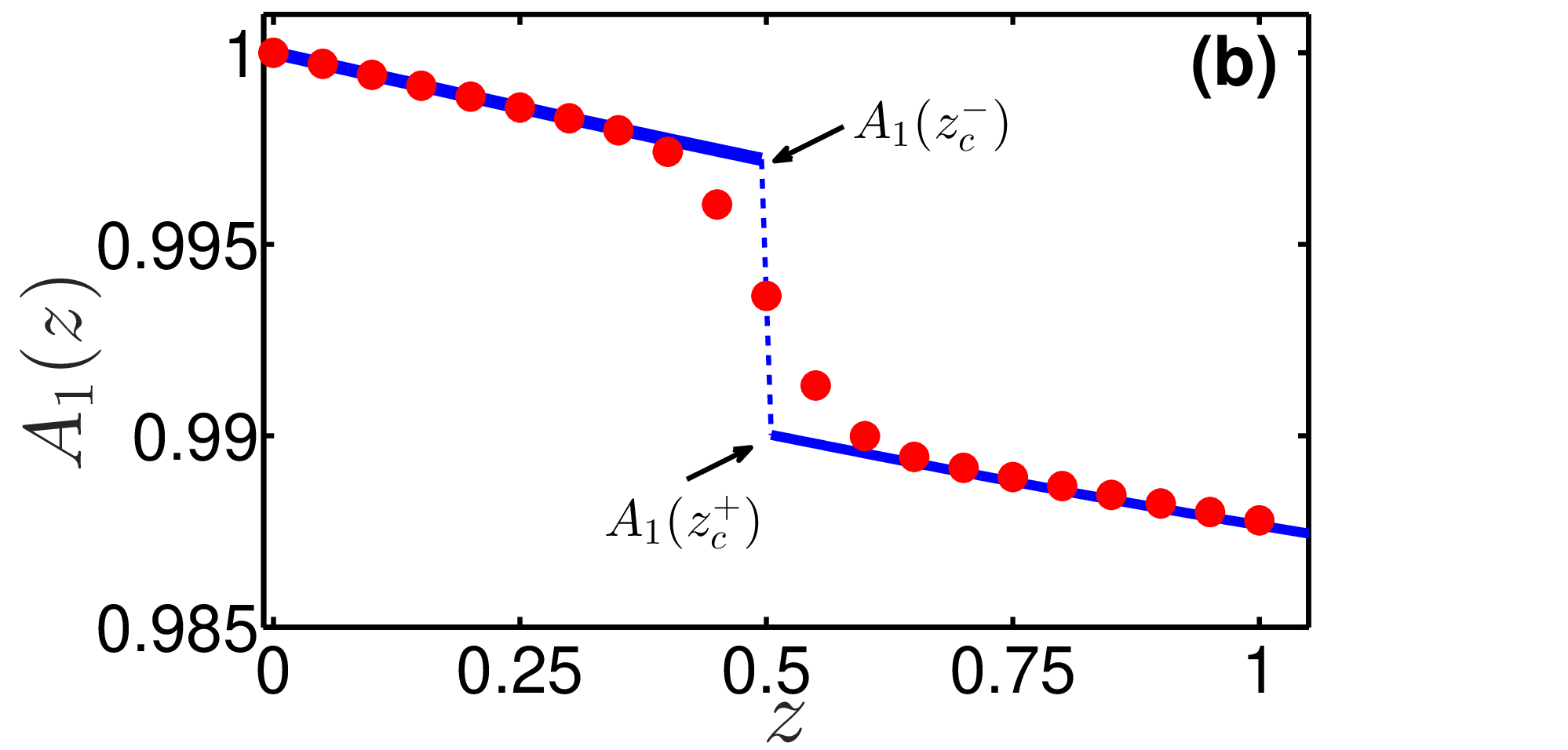} 
\end{tabular}
\caption{(a) Amplitude dynamics of a single single pulse 1 with $\epsilon_{5}=0.01$. The blue solid curve and red circles correspond to $A_{1}(z)$ measured from the theoretical predictions of Eq. (\ref{Amplitude1}) and from numerical simulations of Eq. (\ref{single1}) with $m=2$, respectively. The inset represents the evolution in $z$ of the pulse profile $|\psi_1(t,z)|$. (b) An illustration for the measurement of the collision-induced amplitude shift $\Delta A_{1}^{(c)(num)}$ by Eq. (\ref{simul2}). The two blue solid curves correspond to the approximations of $A_{1}(z)$ for a {\it{single}} pulse without a collision, measured from Eq. (\ref{Amplitude1}) and Eq. (\ref{simul2a}), before and after the collision, respectively, while the red circles correspond to $A_{1}(z)$ from numerical simulations of Eq. (\ref{coll1}) with $m=2$. 
}
\label{fig1}
\end{figure}

\begin{figure}[ptb]
\begin{tabular}{cc}
\epsfxsize=16.5cm  \epsffile{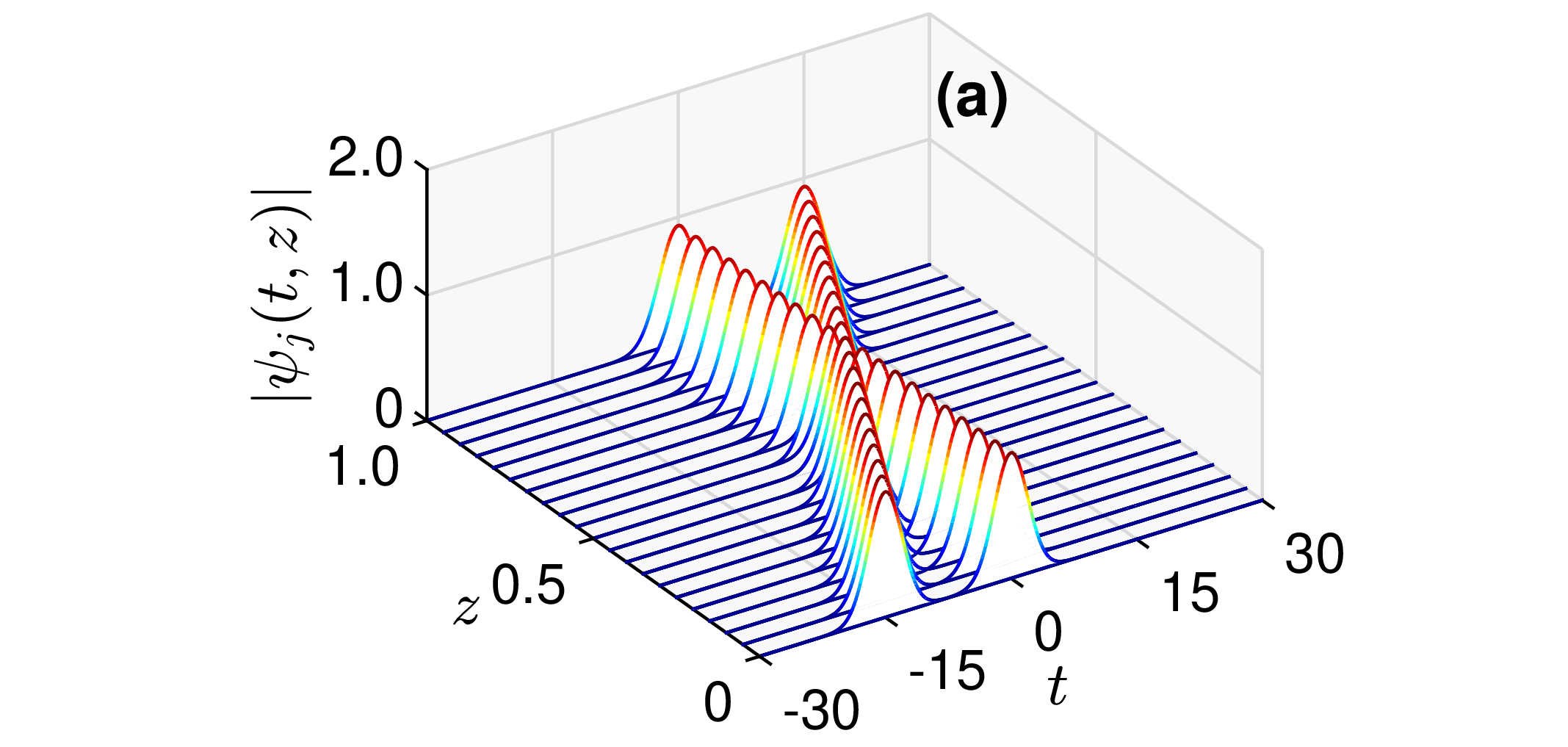} \\
\epsfxsize=15cm  \epsffile{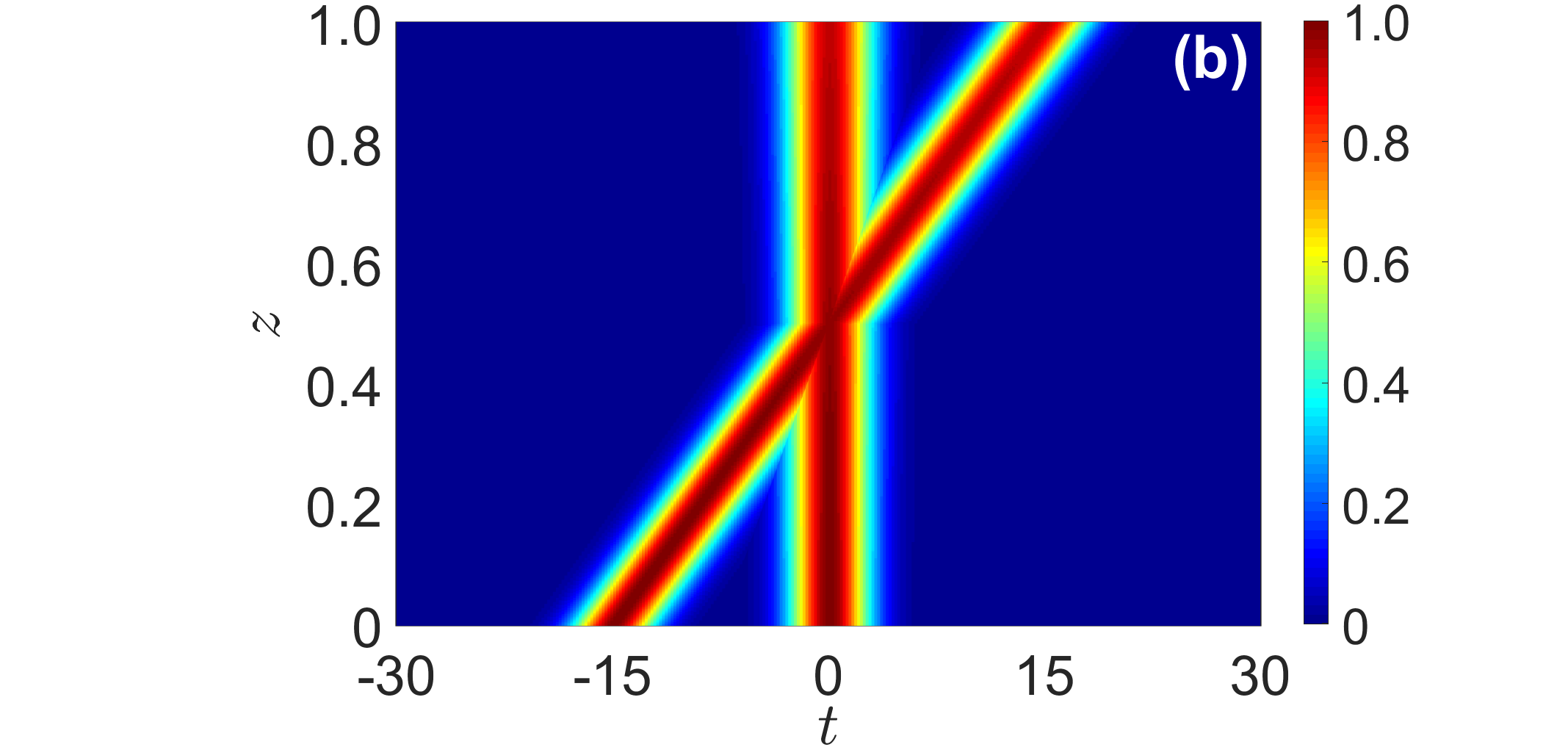} 
\end{tabular}
\caption{A fast collision of two Gaussian pulses at propagation distance $z_{c}=0.5$ in the presence of quintic loss in 3D (a) and its corresponding contour plot (b).  
}
\label{fig2}
\end{figure}

First, we validate Eq. (\ref{Amplitude1}) by the numerical simulations with Eq. (\ref{single1}) for pulse 1 and illustrate an example for calculating $\Delta A_{1}^{(c)(num)}$ in Eq. (\ref{simul2}) with the quintic loss. The parameters used for solving Eqs. (\ref{coll1}) and (\ref{single1}) are $m=2$, $\epsilon_{2m+1}=0.01$, $A_{j0}=1$, $W_{j0}=2$, $\alpha_{j0}=0$, $y_{10}=0$,  $y_{20}=-15$, $\mbox{sgn}(\tilde\beta_{2})=1$, $z_f=1$, and $d_1=30$. Figure \ref{fig1} (a) represents the amplitude dynamics $A_1(z)$ of a single Gaussian pulse in the presence of the quintic loss without any interactions. The inset represents the evolution in $z$ of the pulse profile $|\psi_1(t,z)|$ for $0 \leq z \leq z_f$. The relative error, which is defined by 
$|A_{1}^{(num)}(z)-A_{1}(z)|\times 100/
|A_{1}(z)|$, where $A_{1}^{(num)}(z)$ is measured from Eq. (\ref{single1}), is less than $7.5 \times 10^{-4}$\,\% for $0<z \leq z_f$. This validates the use of Eqs. (\ref{Amplitude1}) and (\ref{simul2a}) to accurately calculate $A_{1}({z_{c}^{-}})$ and $A_{1}({z_{c}^{+}})$ in Eq. (\ref{simul2}). Figure \ref{fig1} (b) captures the amplitude dynamics $A_1(z)$ of pulse 1 in an interaction with pulse 2 at $z_c=0.5$ by numerically solving Eq. (\ref{coll1}) and demonstrates the measurement of the collision-induced amplitude shift $\Delta A_{1}^{(c)(num)}$ by Eq. (\ref{simul2}). As shown in Fig. \ref{fig1} (b), the numerical value of the final amplitude $A_{1}(z_f)$ is a result of the two processes for loss: the self-amplitude shift, which is accurately evaluated by Eqs. (\ref{Amplitude1}) and (\ref{simul2a}), and the rest, which is therefore the collision-induced amplitude shift $\Delta A_{1}^{(c)(num)}$.

Second, we demonstrate the simulations with the following parameters for $m=2$ and $m=3$: $\epsilon_{2m+1}=0.01$, $A_{j0}=1$, $W_{j0}=2$, $\alpha_{j0}=0$, $y_{10}=0$, $y_{20}=\pm 15$,  $\mbox{sgn}(\tilde\beta_{2})=1$, and $8 \leq |d_{1}| \leq 80$. The value of $z_{c}$ is calculated by $z_{c}=|(y_{20}-y_{10})/d_{1}|$. For a fast collision, the final propagation distance $z_f$ can be defined as a distance that the two pulses are well separated after a collision, for example, $z_f \geq 2z_c$. For concreteness, we use $z_f=2z_c$ for all fast collisions. However, we emphasize that another choice of $z_f$ such that $z_f \geq 2z_c$ does not affect the measurement of $\Delta A_{1}^{(c)(num)}$.

Figure \ref{fig2} (a) depicts a particular simulation with Eq. (\ref{coll1}) for a fast collision between two Gaussian pulses in the presence of quintic loss ($m=2$) with $W_{j0}=2$, $d_{1}=30$ and other parameters described as above. Figure \ref{fig2} (b) corresponds to its contour plot. The two pulses collide at the collision distance of $z_c=0.5$. The numerical value for $\Delta A_{1}^{(c)}$ measured from Eq. (\ref{simul2}) is $\Delta A_{1}^{(c)(num)}=-0.007$ while its theoretical prediction calculated from Eq. (\ref{coll11}) is $\Delta A_{1}^{(c)}=-0.0072$. By simulations with different choices of $z_f$, where $z_f \geq 2z_c$, we observe that the measurement of $\Delta A_{1}^{(c)(num)}$ is independent of the choice of $z_f$.

\begin{figure}[ptb]
\begin{tabular}{cc}
\epsfxsize=10cm  \epsffile{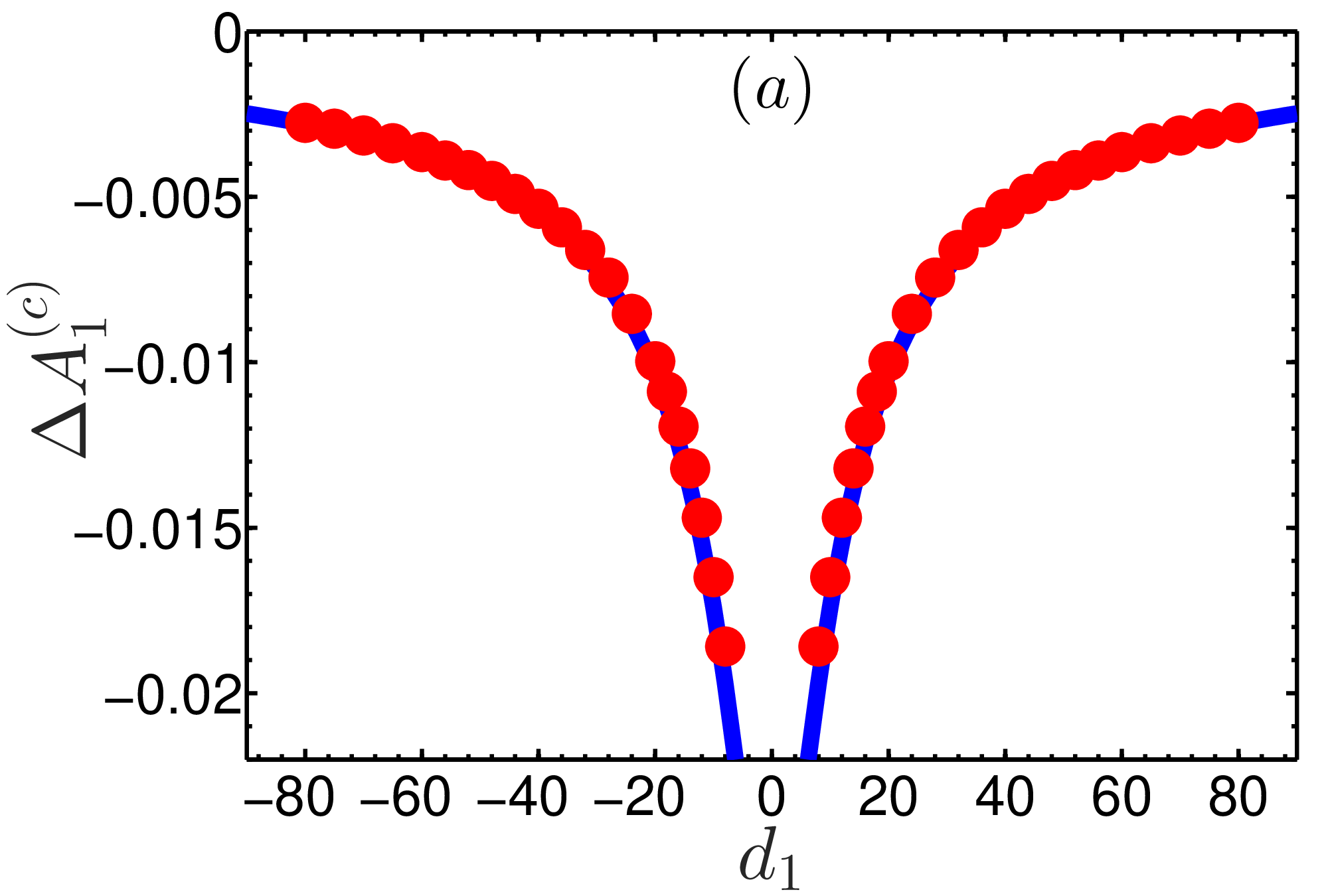} \\
\epsfxsize=10cm  \epsffile{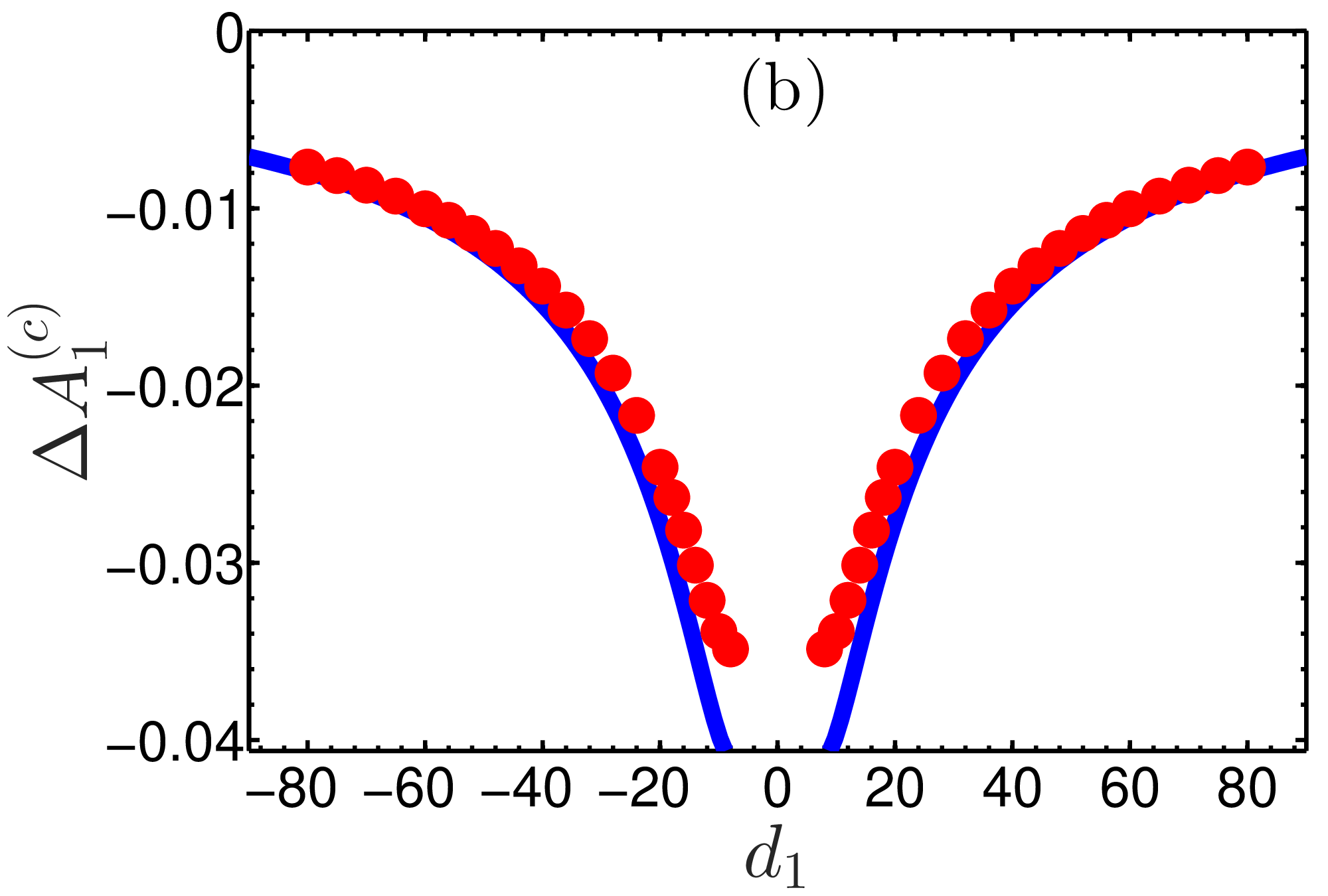} 
\end{tabular}
\caption{Collision-induced amplitude shift in fast collisions of two Gaussian pulses. The blue solid curve and red circles correspond to $\Delta A_{1}^{(c)}$ measured from the theoretical predictions of Eq. (\ref{coll11}) and from numerical simulations of Eq. (\ref{coll1}) with $m=2$ (a) and with $m=3$ (b), respectively. 
}
\label{fig3}
\end{figure}

Figures \ref{fig3}(a) and \ref{fig3}(b) represent the dependence on the collision-induced amplitude shift $\Delta A_1^{\left( c \right)}$ on $d_1$, for $8 \leq |d_{1}| \leq 80$, obtained by simulations with Eq. (\ref{coll1}) with $m=2$ and $m=3$, respectively, along with their analytic predictions of Eq. (\ref{coll11}). As can be seen in Fig. \ref{fig2}, the agreement between the simulations and the analytic predictions are very good. Indeed, for $m=2$, the relative error in the approximation, which is defined by 
$|\Delta A_{1}^{(c)(num)}-\Delta A_{1}^{(c)}|\times 100/
|\Delta A_{1}^{(c)}|$, is less than $4$\% for $|d_{1}| \geq 18$ and less than $2$\% for $|d_{1}| \geq 44$. Even at $|d_{1}| \simeq 8$, the relative error is only $5.7$\%. For $m=3$, the relative error is less than $5$\% for $|d_{1}| \geq 56$ and less than $9.8$\% for $ |d_{1}| \geq 24$. Even at $|d_{1}| \simeq 8$, the relative error is only $15.05$\%. We note that similar results are also obtained for other values of the physical parameters.

Finally, we numerically investigate the collision-induced amplitude dynamics for slow collisions, that is, when the analytic predictions
can be broken-down due to $W_{j0}d_1=\mathcal{O}(1)$. We consider the following parameters for $m=2$ and $m=3$: $\epsilon_{2m+1} = 0.02$, $W_{j0}=1$, $y_{10}=0$, $y_{20}=\pm 5$, $\alpha_{j0}=0$, $z_f=3z_c$, $\mbox{sgn}(\tilde\beta_{2})=1$, and $4 \leq |d_1| <8$. Surprisingly, by varying $d_1$ within $4 \leq |d_1| <8$, we observe that the relative errors are within 5\%-10\% for $m=2$, which is relatively small. For $m=3$, the relative errors are larger, from 15\%-20\%. For instance, for $d_1=5$, the numerical value for $\Delta A_{1}^{(c)}$ measured from Eq. (\ref{simul2}) is $\Delta A_{1}^{(c)(num)}=-0.0297$ while its ``theoretical prediction'' calculated from Eq. (\ref{coll11}) is $\Delta A_{1}^{(c)}=-0.0361$. The relative error is 17.7\%.

In summary, based on the numerical results, we validate our analytic prediction for a collision-induced amplitude shift in fast pulse collisions in the presence of the nonlinear loss.

\section{Conclusions}
\label{concl}
We derived the analytic expression for the collision-induced amplitude shift in a fast collision between two pulses in linear waveguides with the generic weak nonlinear loss. The results revealed that the weak nonlinear loss strongly affects the
collisions of pulses, by causing an additional downshift of pulse amplitudes. More specifically, the collision-induced amplitude shift is of the order of $\epsilon_{2m+1}/|d_{1}|$ in the perturbative calculations, where $0< \epsilon_{2m+1} \ll 1$ and $|d_{1}| \gg 1$. This simple scaling behavior of the order of $\epsilon_{2m+1}/|d_{1}|$ is similar to the scaling for soliton collision-induced amplitude shift found in \cite{PNC2010} for $m=1$ and in \cite{PNG2014} for any $m \geq 1$. This similarity demonstrated that pulses in linear waveguides with the generic weak nonlinear loss exhibit soliton-like behavior in fast two-pulse collisions. Moreover, we showed that the analytic expression for the collision-induced amplitude shift is independent of the pulse shapes of the colliding pulses. The theoretical calculations were confirmed by numerical simulations of the propagation equations in terms of coupled PDEs with $m=2$ and $m=3$ for fast collisions of two Gaussian pulses. Our results, which generalize those in \cite{PNH2017}, provide some
insight into the effects of higher order nonlinear loss on the dynamics of pulses in linear waveguides.

\section*{Acknowledgements}
This research is funded by Vietnam National Foundation for Science and Technology Development (NAFOSTED) under Grant No. 107.99-2019.340. We would like to thank the anonymous Referee(s) for the valuable comments and suggestions.

{}


\begin{thebibliography}{}
\bibitem{Dimensions1} The dimensionless distance $z$ in Eq. (\ref{coll1})  is 
$z=Z/(2L_{D})$, where $Z$ is the dimensional distance, 
$L_{D}=\tau_{0}^{2}/|\tilde\beta_{2}|$ is the dispersion length,
and $\tau_{0}$ is a reference pulse width. 
The dimensionless time is $t=\tau/\tau_{0}$, where $\tau$ is time. 
$\psi_{j}=E_{j}/\sqrt{P_{0}}$, where $E_{j}$ is the 
electric field of the $j$th pulse and $P_{0}$ is peak power.   
$d_{1}=2(\tilde\beta_{12}-\tilde\beta_{11})\tau_{0}/|\tilde\beta_{2}|$, 
where $\tilde\beta_{1j}=n_{gj}/c=1/v_{gj}$, $c$ is the speed of light,  
and $n_{gj}$ and $v_{gj}$ are the group refractive index 
and the group velocity for the $j$th pulse. 
$\epsilon_{1}=2\tau_{0}^{2}\tilde\rho_{1}/|\tilde\beta_{2}|$  
and $\epsilon_{2m+1}=2P_{0}^{m}\tau_{0}^{2}\tilde\rho_{2m+1}/|\tilde\beta_{2}|$, 
where $\tilde\rho_{1}$ and $\tilde\rho_{2m+1}$ are  
the dimensional linear and $(2m+1)-$order of loss coefficients.
\end{thebibliography}
\end{document}